\documentclass[pre,byrevtex,amsmath,amssymb,preprint,showpacs]{revtex4}

\usepackage{graphicx}
\usepackage{dcolumn}
\usepackage{bm}
\begin{document}

\preprint{PREPRINT}

\title[Short Title]{Binary crystals in  two-dimensional two-component Yukawa mixtures}

\author{Lahcen Assoud, Ren\'e Messina, Hartmut L\"owen}
\affiliation
{Institut f\"ur Theoretische Physik II: Weiche Materie,
Heinrich-Heine-Universit\"at D\"usseldorf,
Universit\"atsstrasse 1,
D-40225 D\"usseldorf,
Germany}

\date{\today}

\begin{abstract}

The zero-temperature  phase diagram of binary mixtures of particles
interacting via a screened  Coulomb pair potential
is calculated as a function of composition and charge ratio.
The potential energy obtained by a Lekner summation
 is minimized among  a variety of candidate two-dimensional crystals.
A wealth of different stable crystal structures is identified including 
$A,B,AB_2, A_2B, AB_4$ structures
[$A$ $(B)$ particles correspond to large (small) charge.]
Their elementary cells consist of triangular, square or rhombic
lattices of the $A$ particles with a basis comprising
various structures of $A$ and $B$
particles. For small charge asymmetry there are no intermediate 
crystals besides the pure $A$ and $B$ triangular crystals.
The predicted structures are detectable in experiments on confined
mixtures of charged colloids or dusty plasma sheets.
\end{abstract}
\maketitle

\section{Introduction}
Two-component mixtures in general exhibit much richer crystallization phenomena and polymorphism
than  their one-component counterparts \cite{Tammann} as witnessed by a huge
variety of possible stable binary crystals, e.g. for binary hard sphere systems
\cite{Pronk,Xu,Eldridge1,Bartlett1}. How the  whole crystal phase behavior in mixtures depends on the
interparticle interactions is far from being understood even in equilibrium \cite{Hafner,book}.
This is true also in two spatial dimensions where the number of Bravais lattices
is smaller than in three dimensions. Binary mixtures in two dimensions have been studied
for hard disks \cite{RefDA12} and a complex diagram of close packing was obtained  as a function of their
diameter ratio. More recently, a two-dimensional binary mixture with soft interactions
was considered \cite{Assoud}, namely that for parallel dipoles where the
pair potential scales with the inverse cube
of the interparticle separation. 
A variant of this model has been considered in Ref. \cite{Likos_2007}.
Such systems can be realized in granular matter \cite{Hay}
and in magnetic colloidal suspensions confined to a air-water interface \cite{Maret_1}.
Again, as a function of the ratio of dipole moments of the two species,
a complex stability phase diagram of stable binary crystals was obtained that qualitatively
differs from the hard disk case \cite{RefDA12}. In particular for low asymmetries, the hard disk
system shows a complete separation into pure $A$ and $B$ triangular crystals \cite{RefDA12}
while the soft dipolar systems possesses two  stable mixed crystals as well with
stoechiometric ratio $A_2B$ and $AB_2$ \cite{Assoud}. These differences show that the topology
of the phase diagrams depend on details of the interactions and there is certainly a need
to understand this dependence in more detail.

In this paper, we consider a two-dimensional binary system of Yukawa particles, i.e. the pair
interaction potential $V(r)$  between the particles is a screened Coulomb interaction $\propto \exp ( \kappa r)/r$
where $\kappa$ is the screening constant (or the inverse screening length). This potential interpolates
between the case of hard disks (as obtained in the limit of high $\kappa$) and the unscreened Coulomb case
(as obtained for $\kappa =0$). The latter limit, $V(r) \propto 1/r$ is even softer than the dipolar case where
$V(r) \propto 1/r^3$. The two components are defined by two different charges, i.e.\ different prefactors
in front of the Yukawa interaction. In previous works,
such a classical binary mixture with Yukawa interactions in three-dimensions
has been used as a model to study mixing rules \cite{Rosenfeld}, effective forces \cite{Louis},
fluid-fluid phase separation \cite{phase1,phase2,phase3}, dynamical correlations 
\cite{dynamics1,dynamics2}
and transport properties \cite{Salin}. Likewise the pure (one-component) Yukawa system was also studied
in two-spatial dimensions for fluid structure \cite{Loewen_1992,messina_prl,structure1,structure2},
dynamics \cite{Kalman_PRL_2004,Liu,Libal,Peeters_EPL_2007} and transport properties
\cite{Liu_transport}. Binary mixtures of Yukawa particles in two dimensions have also been studied
for fluid structure \cite{Klein}, adsorption \cite{Gray}, interfaces \cite{Wysocki} and transport \cite{Dzubiella}.
However, the crystallization issue was only addressed in
one-component Yukawa systems (for a recent work, see e.g. \cite{Desgranges})
but never in binary  mixtures.

The Yukawa potential is realized in
{\it charged colloidal suspensions} \cite{Trigger} as well as in  {\it dusty plasmas} \cite{Vaulina},
both for one component systems and mixtures. In fact, highly charged colloidal suspensions
can be confined between highly charged parallel glass plates
\cite{Grier,Brunner,Fontecha} which restricts their motion
practically to two dimensions. The interactions between these macroions are screened
due to the presence of the microscopic microions and additional salt ions.
As in three dimensions, the Debye-H\"uckel screened Coulomb interaction is a reasonable model
for confined charged colloids \cite{Hone,Damico}. Crystallization of binary
charged colloids has been studied experimentally in the bulk. However, a monolayer of a confined
binary mixture of charged colloids has
 not yet been realized
although this is in principle possible as has been shown for sterically-stabilized
\cite{Nugent} and magnetic colloids \cite{Hoffmann2}.
On the other hand, sheets of highly charged dust particles in plasmas
(so-called complex plasmas)  can also be  confined to two dimensions, e.g.\  by
levitating electric fields. The interaction between the dust
particles is again screened such that a Yukawa model is appropriate  \cite{Vaulina,Kong,Hoffmann}.
Highly charged microspheres suspended in a plasma settled in a horizontal monolayer
were studied experimentally and compared to a  two-dimensional Yukawa model \cite{dusty2,dusty3,dusty1}.
There is no principle problem in studying binary mixtures of dust particles but a
concrete realization in an experiments still has to be performed as well.

Apart from its important realizations, our major motivation for our studies
 is to understand the interplay between
the interparticle interaction and the stability of different two-dimensional crystal lattices.
A control of  colloidal composite lattices
may lead to new  photonic crystals \cite{Pine}
to molecular-sieves \cite{Kecht} and to micro- and nano-filters
with desired porosity \cite{Goedel}. The electric properties
of a nanocrystal depend on its superlattice structure \cite{Saunders}.
For these type of applications, it is
crucial  to understand the various stable lattice types  in binary mixtures.

For the two-component two-dimensional Yukawa mixture, we obtain the full phase diagram
at zero-temperature as a function of the charge asymmetry using lattice sums.
As a result, we find a variety of different stable composite lattices.
They include $A,B,AB_2, A_2B, AB_4$  structures.
Their elementary cells consist of (equilateral) triangular, square and  rhombic lattices
of the big particles. These are highly decorated by a basis involving either $A$
particles alone or both $B$ and $A$ particles. The topology of the
resulting phase diagram differs qualitatively from that of hard disk mixtures \cite{RefDA12}
and dipoles \cite{Assoud}.

The paper is organized as follows: In Sec. II  the model is described and possible candidate
structures for crystal lattices in two dimensions are proposed. Results for the phase diagrams
are presented in Sec. III. We conclude finally in Sec. IV.

\section{Model}


%
The model systems used in our study are binary mixtures of
(repulsive) charged  particles made up of
two species denoted as $A$ and $B$. Each component $A$ and $B$  is characterized
by its charge valency  $Z_A$ and $Z_B$, respectively.
These constitutive particles  are confined to a two-dimensional  plane, and interact 
via the Yukawa pair potential. Introducing the ratio $Z=Z_B/Z_A$ 
the pair interaction potentials between two $A$ particles, a $A$- and
$B$-particles, and two $B$-particle at distance $r$ are
%
\begin{eqnarray}
\label{eq_yukawa1}
V_{AA}(r)
\lefteqn{=\kappa V_0 \varphi(r),
\quad V_{AB}(r)=\kappa V_0Z\varphi(r),}\nonumber\\
& & V_{BB}(r)=\kappa V_0Z^2\varphi(r),
\end{eqnarray}
%
respectively. The dimensionless function $\varphi(r)$ is given by
%
\begin{equation}
\label{eq_yukawa2}
\varphi(r)=\frac{\exp(-\kappa r)}{\kappa r},
\end{equation}
%
where the energy amplitude $V_0\kappa$ sets the energy scale.

Our goal is to determine the stable crystalline structures adopted by the system at zero temperature.
We consider a parallelogram as a primitive cell which contains $n_A$ $A$-particles and $n_B$ $B$-particles.
This cell can be described geometrically by the two lattice vectors ${\mathbf a}=a(1,0)$
and ${\mathbf b}=a\gamma(\cos{\theta},\sin{\theta})$, where $\theta$ is the angle between
${\mathbf a}$ and ${\mathbf b}$ and $\gamma$ is the aspect ratio ($\gamma=|{\mathbf b}|/|{\mathbf a}|$).
The position of a particle $i$ (of species $A$)
and  that of a particle $j$ (of species $B$) in the parallelogram is specified by the vectors
${\mathbf r}_{\rm i}^A=(x_i^{A},y_i^{A})$ and
${\mathbf r}_{\rm j}^B=(x_j^{B},y_j^{B})$, respectively.
The total internal energy (per primitive cell) $U$  has the form

\begin{eqnarray}\label{eq_energy}
\lefteqn{U=\frac{1}{2}\sum_{J=A,B}
\sum_{i,j=1}^{n_J}\sideset{}{'}\sum_{\mathbf{R}}V_{JJ}
\left( \left| \mathbf{r}^J_i-\mathbf{r}^J_j+\mathbf{R} \right| \right)}
\nonumber\\
& & + \sum_{i=1}^{n_A}\sum_{j=1}^{n_B}\sum_{{\mathbf R}}
V_{AB}(\left| \mathbf{r}^A_i-\mathbf{r}^B_j+\mathbf{R} \right|),
\end{eqnarray}
%
where ${\mathbf R}=k{\mathbf a}+l{\mathbf b}$ with $k$ and $l$ being integers.
The sums over  ${\mathbf R}$ in Eq. \ref{eq_energy}  run over all lattice cells where the prime indicates
that for  ${\mathbf R=0}$ the terms with $i=j$ are to be omitted.
In order to handle efficiently the long-range nature of the Yukawa interaction at moderate screening
strength, we employed a Lekner-summation (see Appendix A).

We choose to work at prescribed pressure $p$ and zero temperature ($T=0$).
Hence, the corresponding thermodynamic potential is  the Gibbs free energy $G$.
Additionally, we consider interacting particles at composition $X:=n_B/(n_A+n_B)$,
so that the (intensive) Gibbs free energy $g$ per particle
reads: $g=g(p,Z,X)=G/(n_A+n_B)$.
At vanishing temperature,  $g$ is related to the internal
energy per particle $u=U/(n_A+n_B)$ through $ g=u+p/\rho $, where the pressure $p$ is given by
$p=\rho^2(\partial u/\partial\rho)$, and $\rho=(n_A+n_B)/|{\mathbf a} \times {\mathbf b}|$
is the total particle density.
The Gibbs free energy per particle $g$ has been minimized with respect to
$\gamma$, $\theta$  and the position of particles of species $A$ and $B$
within the primitive cell.
In order to decrease the complexity of the energy landscape, we have  limited the number
of variables and considered the following candidates for our binary mixtures:
$A_4B$, $A_3B$, $A_2B$, $A_4B_2$, $A_3B_2$, $AB$,  $A_2B_2$, $A_3B_3$, $A_2B_3$, $AB_2$, $A_2B_4$,
$AB_3$, $AB_4$ and $AB_6$.
For the $AB_6$ and $A_3B_3$ case we have only considered a triangular lattice formed  by the $A$ particles.

%
\begin{figure}
\includegraphics[width=16cm]{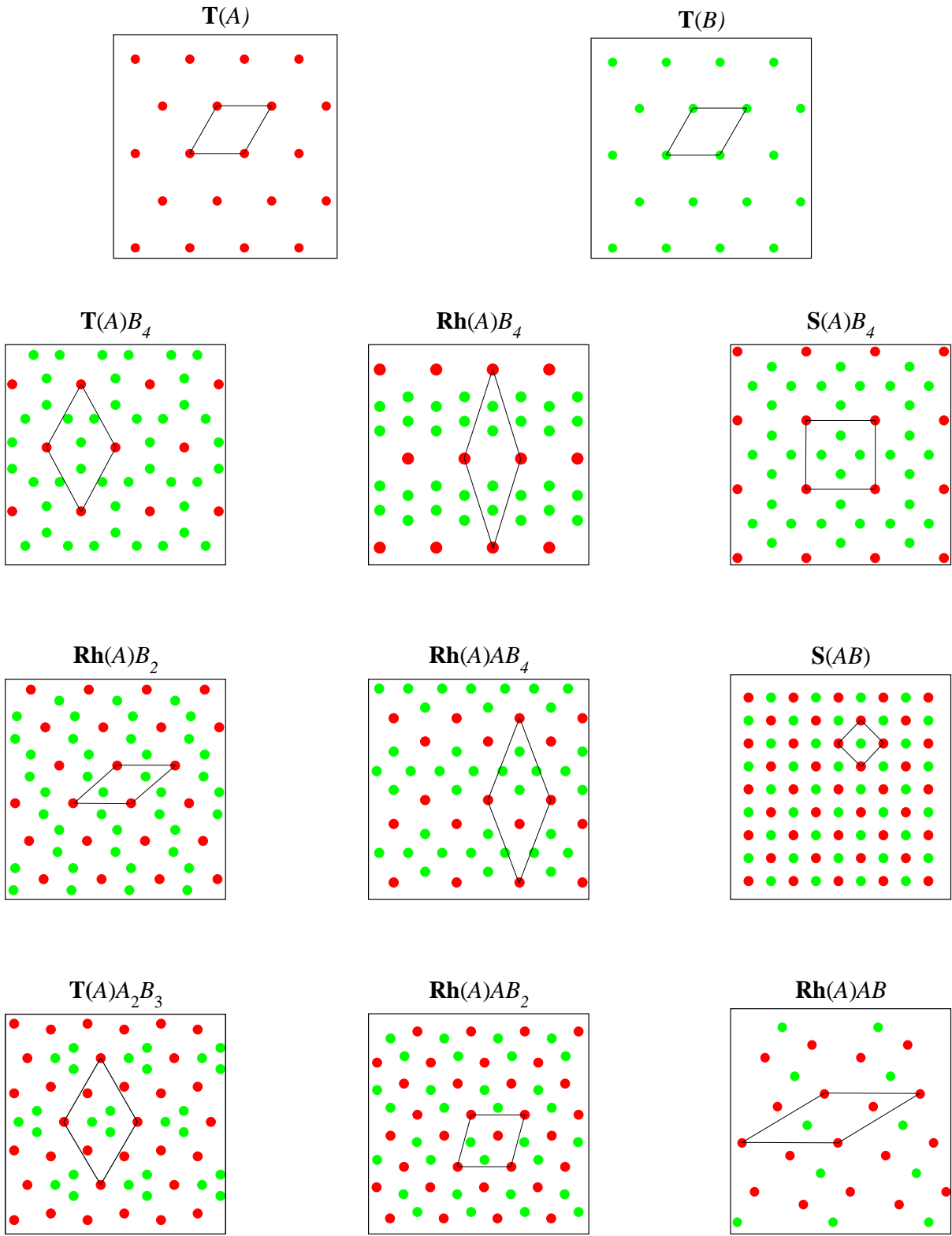}
\caption{The stable binary crystal structures and their primitive cells. The red (green) discs
           correspond to $A$ ($B$) particles.}
\label{fig:bravais_struct}
\end{figure}
%

\section{Results}

\subsection{Phase diagram}

\begin{figure}[]
\includegraphics[width=16cm]{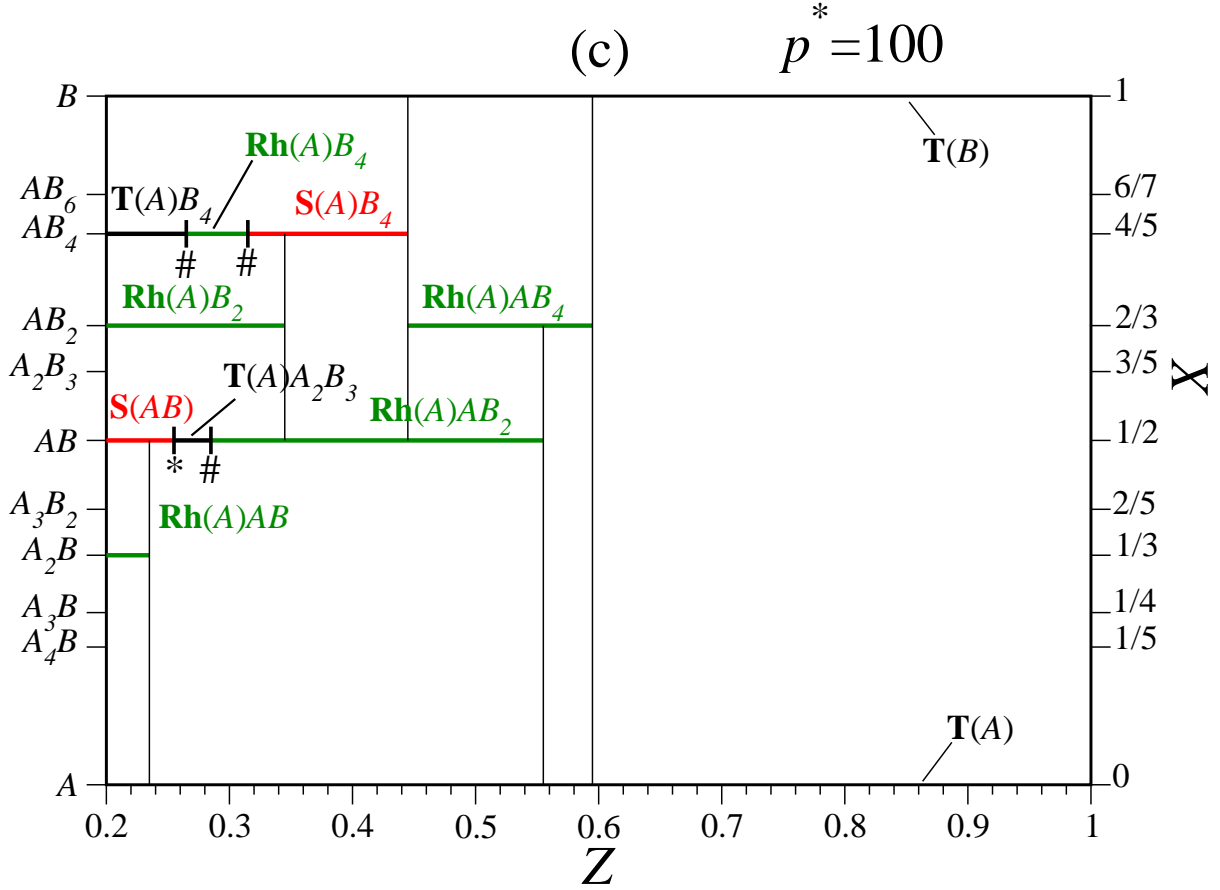}
\caption{The phase diagram in the $(Z,X)$ plane  of charge  asymmetry and composition
    at $T=0$ for a effective pressure (a) $p^*=0.01$, (b) $p^*=1$, (c) $p^*=100$.
    The symbol $\#$ ($\ast$) denote continuous (discontinuous) transitions.}
\label{fig:PD_yuk}
\end{figure}

The ultimate phase diagrams in the $(Z, X)$ plane has been obtained by employing the Maxwell construction.
We recall here that the both dimensionless quantities, namely the charge ratio $Z$ as well as 
the composition $X$, can vary between zero and unity.
A low charge ratio (i.e., $Z$ is close to zero) indicates a strong charge asymmetry, whereas
a high charge ration (i.e., $Z$ is close to unity) represents a large charge symmetry or equivalently
a weak charge  asymmetry.
Given the fact that the phase behavior is getting increasingly complicated upon lowering $Z$, 
involving a huge basket of candidates,
we only present results starting from $Z=0.2$.
Furthermore, in contrast to situations where the pair potential can be described as a power low of
the separation distance (as it was the case in our previous work on dipolar mixtures \cite{Assoud}), 
the phase diagram becomes pressure dependent for Yukawa systems.
To capture this feature, we present results at three well distinct pressures, namely
$p^* \equiv p/ (V_0\kappa^3)=0.01,1$ and 100.
An overview of the resulting stable crystalline phases can be found in Fig. \ref{fig:bravais_struct}.
The corresponding nomenclature of the phase labeling is explained in Table \ref{tab:nom}.
The phase diagrams in the $(Z, X)$ plane for the three reduced pressures 
$p^* =0.01,1$ and 100 are depicted in Fig. \ref{fig:PD_yuk}(a), 
Fig. \ref{fig:PD_yuk}(b), and  Fig. \ref{fig:PD_yuk}(c), respectively.
Note that upon increasing $p^*$ at prescribed $Z$ and $X$ one decreases
the density.

\begin{table}[t]
\caption{
      The stable phases with their Bravais lattice and their basis.}
\begin{tabular}{ll}
\hline
\hline
Phase & Bravais lattice [basis]\\
\hline
{\bf T}($A$)          & Triangular  for $A$ [one $A$ particle]  \\

{\bf T}($B$)          & Triangular  for $B$ [one $B$ particle]  \\

{\bf S}($AB$)         & Square for $A$ and $B$ together [one $A$ and one $B$ particles] \\

{\bf S}$(A)B_n$       & Square for $A$  [one $A$ and $n$ $B$ particles]\\

{\bf Rh}$(A)A_mB_n\quad$   & Rhombic for $A$ [$(m+1)$ $A$ and $n$ $B$ particles] \\

{\bf T}$(A)A_mB_n$    & Triangular for $A$ [$(m+1)$ $A$ and $n$ $B$ particles] \\
\hline
\hline
\label{tab:nom}
\end{tabular}
\end{table}
%

Let us first focus our discussion on the apparently simple phase behavior reported at 
weak charge asymmetry (here roughly $Z \gtrsim 0.5 $, see Fig. \ref{fig:PD_yuk}).
Thereby, the system phase separates into a pure $A$-triangular crystalline phase and a $B$-one
(see also Fig. \ref{fig:bravais_struct}).
This triangular structure obviously corresponds to the single-component ground-state. 
Having in mind that the same phase behavior is reported for hard disks binary mixtures 
at small size asymmetry \cite{RefDA12},
it is meaningful to equally expect a phase separation for moderate or sufficiently large
reduced screening strength $\kappa^* \equiv \kappa / \sqrt \rho$.
For $Z=1$, we have $\kappa^* \approx  3.0, 1.2, 0.4$ for
$p^* =0.01,1,100$, respectively, so that the phase separation is certainly
to be expected for moderate pressures (here $p^*=0.01$ and possibly  $p^* =1$) 
when referring to the hard disk limit \cite{RefDA12}.   
%

%
\begin{figure*}[]
\includegraphics[width=16cm]{fig2a.eps}
%
\includegraphics[width=16cm]{fig2b.eps}
\end{figure*}
%

What is now less obvious, still in the regime of weak charge asymmetry, 
is the phase separation reported in Fig. \ref{fig:PD_yuk}(c)
for $p^* =100$, where  $\kappa^* \approx 0.4$ for large charge symmetry.
Recently, we have shown for dipolar binary mixtures \cite{Assoud}, 
whose pair potential is governed by $1/r^3$,
that, at weak dipolar asymmetry 
(the analogous quantity to the charge ratio in our present study),    
the {\it stable mixtures} $A_2B$ and  $AB_2$ (who are globally triangular) set in.
This phase behavior contrasts therefore strongly with that 
presently reported for Yukawa mixtures, see Fig. \ref{fig:PD_yuk}(c).
Given the fact that at weak screening the Yukawa pair potential is
well approximated by a $1/r$ dependence, which is even softer than 
$1/r^3$, it is legitimate to expect stable mixtures in the regime of
weak screening and charge asymmetry.
In order to check this idea we have performed additional calculations at 
$p^* =10^{10}$ with $Z=0.99$ leading to reduced screening strengths
of the order of $10^{-2}$. 
Those values for $ \kappa^*$ turn out to be still too large to recover
the phase behavior found at $1/r^3$-pair interactions \cite{Assoud}.    
The consideration of even much smaller screening strengths 
(say roughly of the order of $10^{-7}$)  are numerically not 
tractable within reasonable CPU time. 
Unfortunately, the implementation of a 
direct Lekner and/or Ewald sum for the $1/r$-pair interactions 
is delicate at {\it prescribed pressure}, since the lack of electroneutrality
involves the presence of an artificial homogeneous neutralizing background which is 
thermodynamically only consistent at  {\it prescribed density} \cite{Bonsall_PRB_1977}.
Consequently, although we have a strong intuition about the stability 
of mixtures at weak charge asymmetry and screening, we can not prove 
it here on computational basis.   

We now briefly address the more complicated phase behavior reported at 
strong charge asymmetry, see Fig. \ref{fig:PD_yuk} with $Z \lesssim 0.5$. 
As a clear general trend, it is found that the number of stable phases
increases with growing pressure. This feature is in agreement with the idea that
mixing is favored upon softening the pair potential. 

A common and remarkable feature in this regime of strong charge asymmetry 
(see Fig. \ref{fig:PD_yuk}) is the imposing stability of the composition $X=1/2$. 
This feature was also reported for dipolar mixtures \cite{Assoud}.
More specifically, the following cascade 
${\bf S}(AB) \to {\bf T}(A)A_2B_3 \to {\bf Rh}(A)AB_2$ is found upon increasing
$Z$, see Fig. \ref{fig:PD_yuk} and Fig. \ref{fig:bravais_struct} for the corresponding structures. 
Thereby, the transition ${\bf S}(AB) \to {\bf T}(A)A_2B_3$
is discontinuous whereas ${\bf T}(A)A_2B_3 \to {\bf Rh}(A)AB_2$ is continuous,
see Fig. \ref{fig:PD_yuk}. Note that, for $p^*=0.01$ shown in Fig. \ref{fig:PD_yuk}(a), 
the stability of the square phase
${\bf S}(AB)$ occurs for values of $Z$ smaller than 0.2 that are not shown here.    

\subsection{Thermodynamical properties}

\subsubsection{Constant pressure}

In this part, we investigate some thermodynamic properties
such as the reduced density $\rho^*$ or the reduced Gibbs free energy
$g^* \equiv g/(V_0 \kappa )$, as obtained prior the Maxwell construction.
Although the pressure considered here is fixed at $p^*=100$, 
very similar results are obtained for the two other pressures.

\begin{figure}[]
\includegraphics[width=9.0cm]{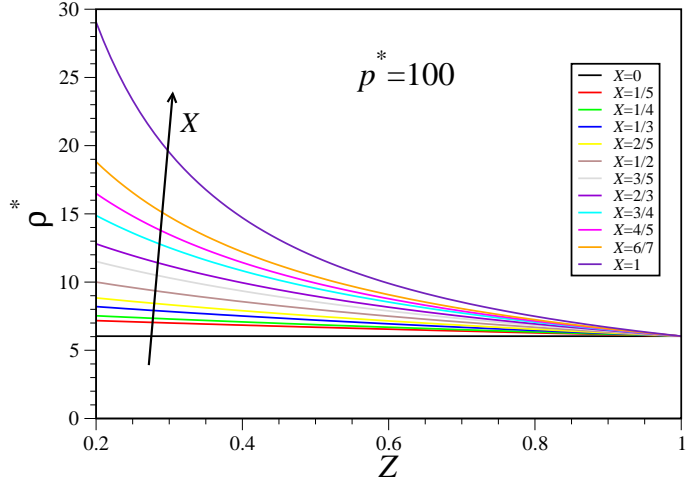}
\caption{Reduced density $\rho^*$ (prior the Maxwell construction) as a function 
         of the charge ratio $Z$ for various compositions $X$ at prescribed reduced pressure $p^*=100$.
         The arrow indicates growing $X$.}
\label{fig:density_at_high_p}
\end{figure}

The reduced density $\rho^*$ as a function of the charge ratio $Z$
at different compositions $X$ is sketched in Fig. \ref{fig:density_at_high_p}. 
At given composition $X$, the density decreases monotonically with $Z$,
see Fig. \ref{fig:density_at_high_p}.
This effect can be simply explained as follows: Upon increasing $Z$ the repulsive
$A-B$ and $B-B$ pair interactions increase accordingly, so that to keep the pressure fixed
the system has to decrease its density.   
Moreover, at prescribed charge ratio, Fig.  \ref{fig:density_at_high_p} indicates
that the density   increases with the composition.
This feature can also be explained with simple physics:
Upon enlarging the composition $X$, the proportion of {\it weakly} charged $B$-particles
increases accordingly, so that to keep the pressure constant the system has to
increase its density. 

\begin{figure}[b]
\includegraphics[width=9.0cm]{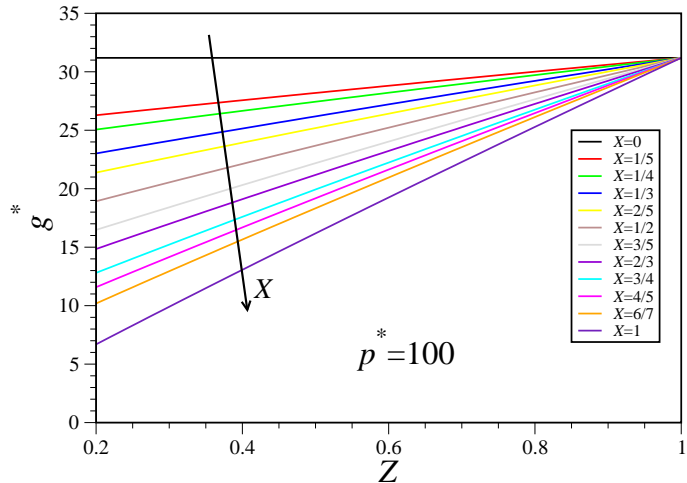}
\caption{Same as Fig. \ref{fig:density_at_high_p} but for $g^*$.}
\label{fig:g_at_high_p}
\end{figure}

The reduced Gibbs free energy  $g^*$ as a function of the charge ratio $Z$
at different compositions $X$ is sketched in Fig. \ref{fig:g_at_high_p}. 
Recalling that $g^* = u^* + p^*/\rho^*$ [with $u^* \equiv  u/(V_0 \kappa)$],
the behavior of $g^*$ exhibited in Fig. \ref{fig:g_at_high_p} can be
equally well rationalized when advocating the just explained 
behavior of $\rho^*$ described in Fig. \ref{fig:density_at_high_p}.
Since the reduced internal energy $u^*$ decreases with growing $X$ 
and therefore with growing $\rho^*$
at prescribed $Z$, it is clear that $g^*$ decreases with growing $X$ at given 
$Z$, as seen in Fig. \ref{fig:g_at_high_p}.  
Besides, at given composition $X$, Fig. \ref{fig:g_at_high_p} shows
that $g^*$ increase with growing $Z$, as expected.

\subsubsection{Constant composition}
\begin{figure}[]
\includegraphics[width=9.0cm]{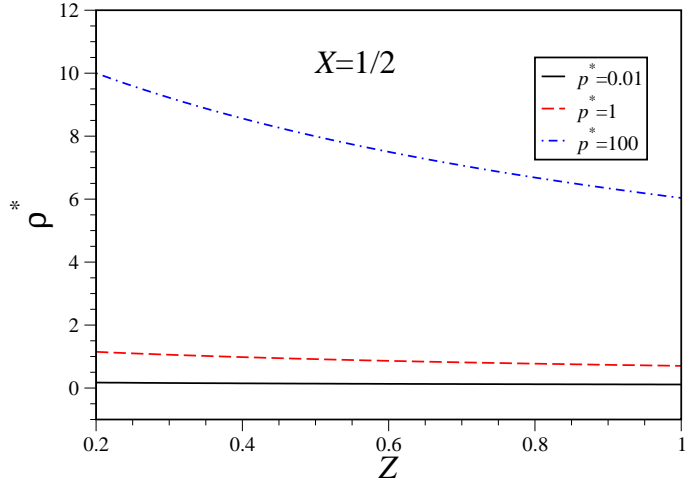}
\caption{Reduced density $\rho^*$ (prior the Maxwell construction) as a function of the charge ratio $Z$ 
         for various reduced pressures $p^*$ at prescribed composition $X=1/2$.}
\label{fig:density_at_5050}
\end{figure}

\begin{figure}
\includegraphics[width=9.0cm]{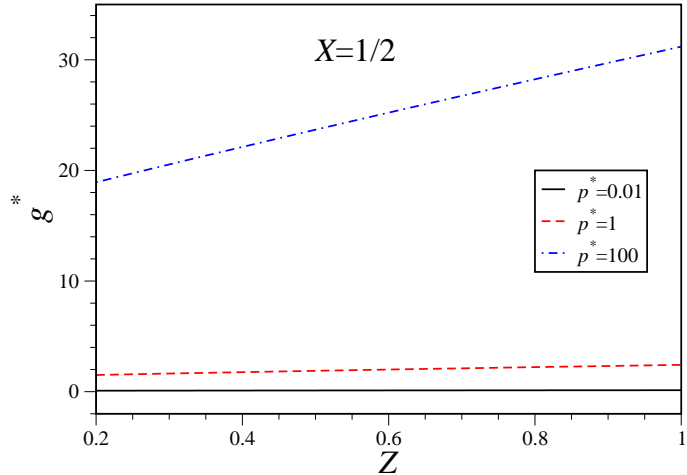}
\caption{Same as Fig. \ref{fig:density_at_5050} but for $g^*$.}
\label{fig:g_at_5050}
\end{figure}

We now analyze $\rho^*$ and  $g^*$ at $X=1/2$ as a function of $Z$
for different values of $p^*$. As far as the behavior of $\rho^*$ is concerned,
the new information provided by Fig. \ref{fig:density_at_5050} is that
$\rho^*$ increases with growing pressure at given charge ratio, as it should be.   
The same qualitative feature is also observed for $g^*$ in Fig. \ref{fig:g_at_5050}.
A closer inspection of Fig. \ref{fig:density_at_5050} and Fig. \ref{fig:g_at_5050} 
suggests that $\rho^*$ and  $g^*$ increase rather slowly with $p^*$ (at given $Z^*$).

\section{Concluding remarks}

In conclusion we have determined the ground-state (i.e. zero-temperature)
phase diagram for a two-component Yukawa monolayer
at various pressure for arbitrary compositions and a broad range of charge
asymmetries. Among a big number of candidate phases, a
wealth of different composite lattices has been found to be stable.
The larger the charge asymmetry, the more complex is the phase diagram.
At low asymmetry the system shows demixing into pure $A$ and $B$
crystals similar to hard disks but different from the soft
inverse cube interaction valid for dipoles.
The results are in principle detectable in binary mixtures of
charged colloids confined between two charged plates or levitated dusty
plasma sheets.

It would be interesting to study the effect of finite temperature.
We expect that the topology of the phase diagram does not change upon
gently increasing the temperature though this could change close to melting.
 When cooling a two-component fluid down,
glass formation in the binary systems at finite temperature
may be a fascinating topic as well \cite{Hamanaka_PRE_2006} to be studied in the future.
In fact, it has been speculated that the underlying crystallization
into the stable
crystal lattices may control vitrification \cite{Tanaka_PRL_2007}
 and therefore are findings are
relevant for the structure of glasses as well.

\acknowledgments

We thank T. Palberg and H. Tanaka for helpful discussions.
This work was supported by the DFG via the SFB TR6 (project section D1).

\appendix
\section{Lekner sums for Yukawa interactions in two dimensional systems}
We consider a primitive cell in the shape of a parallelogram, which contains a set of  $n=n_A+n_B$ particles 
interacting via Yukawa potentials. The parallelogram repeated in the $xy$ plane gives a 2-dimensional lattice, 
and can be described by two lattice vectors ${\mathbf a}=(a_x,0)$ and  ${\mathbf b}=(b_x,b_y)$.
In the parallelogram, the position of a charge valency $Z_i$ is defined by ${\mathbf r}_i=(x_i,y_i)$.

The total interaction energy per cell is given by
%
\begin{equation}
\label{app1}
\frac{U}{V_0}=\frac12\sum_{i=1}^{n}\sum_{j\not=1}^{n}Z_iZ_j
\Phi({\mathbf r}_{ij})+\frac12\sum_{i=1}^{n}Z_i^2\Phi_0
\end{equation}   
%
with   
%
\begin{equation}
\label{app2}
\Phi({\mathbf r})=\sum_{\mathbf R}\frac{\exp(-\kappa|{\mathbf r}+{\mathbf R}|)}{|{\mathbf r}+{\mathbf R}|}
\quad \mbox{and} \quad \Phi_0=\sum_{\mathbf R\not=0}\frac{\exp(-\kappa|{\mathbf R}|)}{|{\mathbf R}|},
\end{equation}   
%
where
%
\begin{equation*}
|{\mathbf r}+{\mathbf R}|=\sqrt{(x+a_xl+b_xm)^2+(y+b_ym)^2} \quad \mbox{and} \quad 
|{\mathbf R}|=\sqrt{(a_xl+b_xm)^2+(b_ym)^2}.
\end{equation*}   
%
Here ${\mathbf R}=l{\mathbf a} +m{\mathbf b} $ with $l$ and $m$ being integers.
The slowly convergent sums over lattice sites (Eq. \ref{app2}) can not be efficiently 
used in a numerical calculation, so that we will transform them into rapidly convergent 
forms using a Lekner Method \cite{Lekner,Mazars_yu_lekner}.  
With the help of the following integral representation
%
\begin{equation}
\label{app3}
\frac{\exp(-\kappa|{\mathbf r}+{\mathbf R}|)}{|{\mathbf r}+{\mathbf R}|}=
\frac{1}{\sqrt{\pi}}\int_0^{\infty}\frac{dt}{\sqrt{t}}\exp{(-\frac{\kappa^2}{4t}-|{\mathbf r}+{\mathbf R}|^2t)},  
\end{equation}   
%
we obtain
%
\begin{equation}
\label{app4}
\Phi({\mathbf r})= \frac{1}{\sqrt{\pi}}\int_0^{\infty}\frac{dt}{\sqrt t}\left\{
\exp(-\frac{\kappa^2}{4t})\sum_{m=-\infty}^{\infty}\sum_{l=-\infty}^{\infty}
\exp \left[-(y+mb_y)^2t\right]
\exp \left[-\left(\frac{x}{a_x}+l+m\frac{b_x}{a_x}\right)^2a_x^2t\right]\right\}.
\end{equation}
Now, to get further, we apply a 1-dimensional Poisson summation
\begin{equation}\label{app5}
\sum_{l=-\infty}^{\infty}\exp{\left[-(\alpha+\beta l)^2t\right]}=\frac{\sqrt{\pi}}{\beta\sqrt{t}}
\sum_{k=-\infty}^{\infty}\exp{\left(i2\pi k\frac{\alpha}{\beta}\right)}
\exp{\left( -\frac{\pi^2k^2}{\beta^2}\frac1t\right)},
\end{equation}
%
which provides
%
\begin{equation}
\label{app6}
\sum_{l=-\infty}^{+\infty}\exp \left[-(\frac{x}{a_x}+l+m\frac{b_x}{a_x})^2a_x^2t)\right]
=\frac{1}{|a_x|}\sqrt{\frac{\pi}{t}}\left[1+2\sum_{k=1}^{+\infty}
\cos\left[ 2\pi k\left(\frac{x}{a_x}+m\frac{b_x}{a_x}\right)\right]\exp\left(-\pi^2 k^2/a_x^2t\right)\right].
\end{equation}
%
Inserting Eq. (\ref{app6}) into Eq. (\ref{app4}) yields:
%
\begin{eqnarray}
\label{app7}
\Phi({\mathbf r})&=&\frac{1}{|a_x|}\sum_{m=-\infty}^{\infty}\int_0^{\infty}\frac{dt}{t}
\exp\left[-\frac{\kappa^2}{4t}-(y+mb_y)^2t\right]\nonumber\\
&&+\frac{2}{|a_x|}\sum_{k=1}^{+\infty}\sum_{m=-\infty}^{+\infty}
\cos\left[2\pi k\left(\frac{x}{a_x}+m\frac{b_x}{a_x}\right)\right]\nonumber\\
&&\times\int_0^{\infty}\frac{dt}{t}
\exp\left[-\left(\kappa^2+\frac{4\pi^2k^2}{a_x^2}\right)\frac{1}{4t}-(y+mb_y)^2t\right]
\end{eqnarray}
%
Now, taking into account the following relation
%
\begin{equation}
\label{app8}
\int_0^{\infty}\frac{dt}{t}\exp{\left(-\frac{B^2}{4t}-C^2t\right)}=2K_0(BC)
\end{equation}
%
where $K_0$ is the zeroth order modified Bessel function of the second kind, 
The final expression for ${\Phi(\mathbf r})$ reads:
%
\begin{eqnarray}
\label{app9}
\Phi({\mathbf r})&=&\frac{2}{|a_x|}\sum_{m=-\infty}^{+\infty}K_0\left(\kappa|y+mb_y|\right)\nonumber\\
&&+\frac{4}{|a_x|}\sum_{k=1}^{\infty}\sum_{m=-\infty}^{+\infty}
\cos\left[2\pi\left(\frac{x}{a_x}+m\frac{bx}{a_x}\right)\right]\nonumber\\ 
&&\times K_0\left[|y+mb_y|\sqrt{\kappa^2+\frac{4\pi^2k^2}{a_x^2}}\right]
\qquad\qquad \mbox{for } y\not=0
\end{eqnarray}
%
and the ``self'' contribution $\Phi_0$ 
%
\begin{eqnarray}
\label{app10}
\Phi_0&=&\frac{4}{|a_x|}\sum_{m=1}^{\infty}K_0(\kappa mb_y)\nonumber\\
&&+\frac{8}{|a_x|}\sum_{k=1}^{\infty}\sum_{m=1}^{\infty}\cos \left(2\pi km\frac{bx}{a_x}
\right)K_0\left[mb_y\sqrt{\kappa^2+\frac{4\pi^2k^2}{a_x^2}}\right]\nonumber\\
&&-\frac{2}{|a_x|}\ln \left[1-\exp(-\kappa a_x)\right]
\end{eqnarray}
%
In the limit of a rectangular based cell, i.e setting $b_x=0$, 
one obtains the formulas for the cross and self-energies
that are identical to those derived in \cite{Mazars_yu_lekner} with $z=0$.



\end{document}